# Novel models for fatigue life prediction under wideband random loads based on machine learning

Hong Sun[1], Yuanying Qiu [1, *], Jing Li [1, *], Jin Bai [2], Ming Peng [3]

[1] School of Mechatronic Engineering, Xidian University, No. 2 South Taibai Road, Xi'an, China, 710071.

[2] Xi'an Aerospace Propulsion Test Technology Institude, Xi'an, China, 710100.

[3] Hunan Province Motor Vehicle Technician College, Shaoyang, China, 422001.

* Corresponding author.

E-mail addresses: yyqiu@mail.xidian.edu.cn (Yuanying Qiu),

lijing02010303@163.com (Jing Li).

**Abstract**：Machine learning as a data-driven solution has been widely applied in the field of fatigue lifetime prediction. In this paper, three models for wideband fatigue life prediction are built based on three machine learning models, i.e. support vector machine (SVM), Gaussian process regression (GPR) and artificial neural network (ANN). The generalization ability of the models is enhanced by employing numerous power spectra samples with different bandwidth parameters and a variety of material properties related to fatigue life. Sufficient Monte Carlo numerical simulations demonstrate that the newly developed machine learning models are superior to the traditional frequency-domain models in terms of life prediction accuracy and the ANN model has the best overall performance among the three developed machine learning models.

## 1.Introduction

In reality, there are many engineering structures subjected to random loads arising from wind, rough pavements and waves [1]. The power spectra of the structural response can usually be determined by spectral analysis. Once the power spectra is determined, as illustrated in Fig. 1, there exist generally two approaches [2] for conducting random fatigue life prediction, i.e. the time-domain fatigue analysis (TDFA) and the frequency-domain fatigue analysis (FDFA).

As for TDFA, firstly the response power spectra is converted into stress time-domain signals through Monte Carlo numerical simulation, then many stress rainflow cycles are obtained from these signals using the rainflow counting [3], and finally the estimation of random fatigue life is achieved by using the amplitudes of stress cycles, material properties and Miner's rule [4]. With regard to FDFA, every frequency-domain model describing the intricate relation between the power spectra of a random load and the fatigue damage can be used directly to predict random fatigue life, omitting the rainflow counting and numerical simulation [2,5-9].

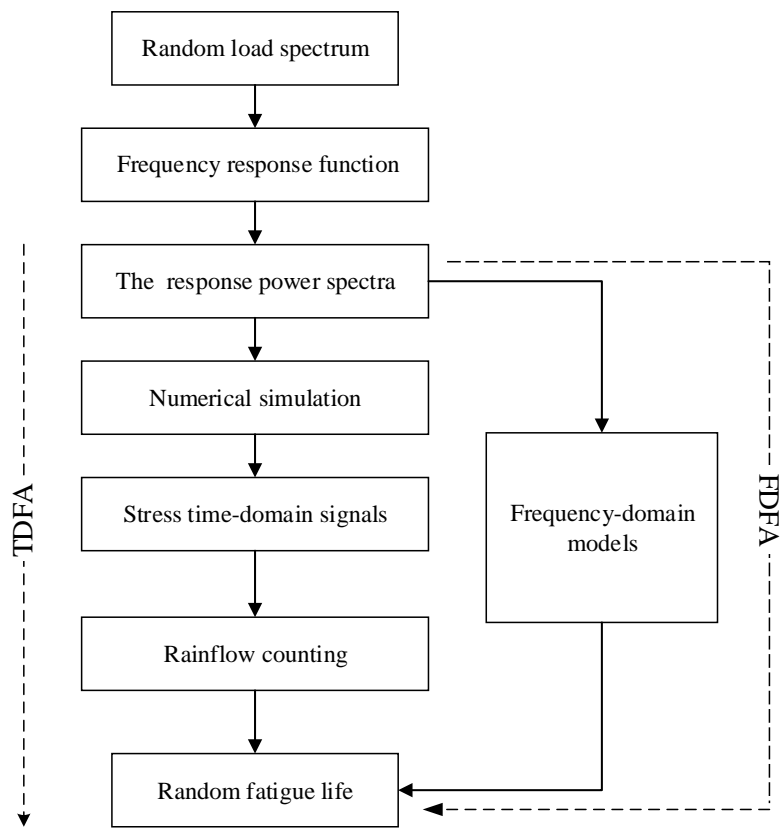

Fig. 1 The flowchart of two types of analysis methods for predicting random fatigue life

While acknowledged as the more precise technique for predicting random fatigue life, TDFA demands significant computational resources [10,11]. Thus, various frequency-domain models [2,5-9] have been devised to enhance computational efficiency. Remarkably, FDFA using these models achieve a prediction accuracy that is nearly on par with TDFA using the rainflow counting method.

Frequency-domain models can be categorized into narrowband and wideband models. For the former, Bendat [7] has formulated the famous narrowband approximation formula, but the result may be rather conservative if this formula is used to estimate the fatigue life of an engineering structure under a wideband random load whose Vanmarcke bandwidth parameter is 0.1 to 0.95 [12]. Therefore, some researchers have developed several wideband models for predicting the wideband random fatigue life [2,5,6,8]. Wirsching and Light [8] adopted a correction factor to modify the wideband random fatigue life calculated by the narrowband approximation formula. Wu et al. [13] developed a wideband model by modifying the narrowband formula and using width parameters as weight coefficients. By executing Monte Carlo numerical simulations with 70 different power spectra samples based on two kinds of power spectra patterns in Fig. 2, Dirlik [6] developed an empirical formula combing the Rayleigh distribution function with the exponential function to approximate the rainflow amplitude probability density function for random fatigue life prediction. In place of the exponential function, the empirical formula developed by Zhao and Baker [2] combines the Weibull and Rayleigh distribution functions. In addition, Tovo and Benasciutti [5] developed the Tovo–Benasciutti (TB) model by performing Monte Carlo numerical simulations with 286 different power spectra samples. Wu et al. [14] derived an empirical formula for predicting the wideband random fatigue life of offshore structures. Although the above models can predict fatigue life well, only a small number of power spectra samples were used in their development, limiting their generalizability [12,15]. Instead, with numerous power spectra samples as training dataset, the frequency-domain models constructed via machine learning may have greater generalizability.

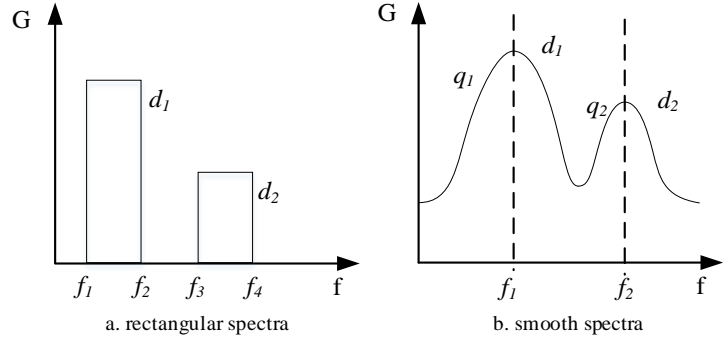

Fig. 2 Two power spectra patterns used by Dirlik

At present, machine learning models have been applied extensively in the field of fatigue life prediction. Kang et al. [16] developed an ANN model to fit the complex relationship between the 0th order spectral moment and shape parameters of a power spectra and the rainflow amplitude probability function. Durodola et al. [15] estimated wideband random fatigue life using an ANN model with the spectral moments of a power spectra and material parameters as inputs and the fatigue damage as an output. Moises et al. [17] employed an ANN model to analyze the fatigue life of chassis components. He et al. [18] have applied a machine learning model to the prediction of fatigue life of metallic materials. Zhang et al. [19] employed traditional machine learning models for the fatigue life prediction of the austenitic stainless steel. Horňas et al. [20] devised three machine learning models to predict the fatigue life of additively manufactured Ti-6Al-4V.

From the above description, machine learning models have a powerful potential in fatigue life prediction. Therefore, in this study, three classical machine learning models (i.e. SVM, GPR and ANN models) are used to perform wideband random fatigue life prediction. The framework of this paper is as follows: Firstly, some related theories of random process and two types of fatigue analysis methods (i.e. TDFA and FDFA) are introduced in Section 2. Then, the specific implementation procedure of the three machine learning models proposed in this paper is described in Section 3, involving the selection of models' inputs and output, the establishment of a training database, and the relevant parameter settings of machine learning models. Finally, in Section 0, the three machine learning models developed in this paper and two traditional frequency-domain models (i.e. TB and Dirlik models) are compared in terms of prediction accuracy and computational efficiency.

## 2. Theoretical background

### 2.1 Properties of stochastic processes

For a stationary Gaussian process $\{X(t)\}$ with a mean of zero, the spectral moments reveal the intrinsic time-domain characteristics of the process itself and are defined as [21]:

$$\lambda_m = \int_0^\infty \omega^m G_X(\omega) d\omega \quad m = 1, 2, \ldots \tag{1}$$

where $\omega$ is the angular frequency and $G_X(\omega)$ represents one-sided power spectra.

There exist several specific relations between spectral moments and variances of a random process $\{X(t)\}$ and its derivatives $\{\dot{X}(t)\}$ and $\{\ddot{X}(t)\}$:

$$\lambda_0 = \sigma_X^2, \quad \lambda_2 = \sigma_{\dot{X}}^2, \quad \lambda_4 = \sigma_{\ddot{X}}^2 \tag{2}$$

in which $\lambda_0$, $\lambda_2$ and $\lambda_4$ are the 0th order, 2nd order and 4th order spectral moments respectively; $\sigma_X^2$, $\sigma_{\dot{X}}^2$, and $\sigma_{\ddot{X}}^2$ denote the variances of $\{X(t)\}$, $\{\dot{X}(t)\}$ and $\{\ddot{X}(t)\}$ respectively.

Moreover, the expected up-crossing rate $v_0$ and the peak rate $v_P$ of the process $\{X(t)\}$ are expressed by the spectral moments, respectively [5]:

$$v_0 = \frac{1}{2\pi}\sqrt{\frac{\lambda_2}{\lambda_0}}; \quad v_P = \frac{1}{2\pi}\sqrt{\frac{\lambda_4}{\lambda_2}} \tag{3}$$

And some commonly-used bandwidth parameters can be calculated by spectral moments as follows:

$$\alpha_{0.75} = \frac{\lambda_{0.75}}{\sqrt{\lambda_0 \lambda_{1.5}}}, \quad \alpha_1 = \frac{\lambda_1}{\sqrt{\lambda_0 \lambda_2}}, \quad \alpha_2 = \frac{\lambda_2}{\sqrt{\lambda_0 \lambda_4}} \tag{4}$$

Besides, the Vanmarcke bandwidth parameter [22] $q_x$ is also frequently used:

$$q_x = \sqrt{1 - \frac{\lambda_1^2}{\lambda_0 \lambda_2}} = \sqrt{1 - \alpha_1^2} \quad 0 \leq q_x \leq 1 \tag{5}$$

Notably, the closer the $q_x$ of a random process is to 1, the closer the process is to wideband.

**2.2 Time-domain fatigue analysis**

In the time domain, the fatigue damage per unit time $\bar{D}_{\text{RFC}}$ is usually calculated by combining the rainflow counting and Miner's rule [23]. $\bar{D}_{\text{RFC}}$ can be expressed as [23]:

$$\bar{D}_{\text{RFC}} = \frac{1}{T}\left(\sum_{i=1}^{N_\text{T}} \frac{S_i^k}{C}\right) = \frac{1}{TC}\left(\sum_{i=1}^{N_\text{T}} S_i^k\right) \tag{6}$$

where $C$ and $k$ are the parameters of the equation $NS^k = C$ for *S-N* curve respectively, $S_i$ is the stress amplitude of the *i*th rainflow cycle, and $N_\text{T}$ is the number of rainflow cycles during time length $T$ of a time-domain signal sample.

**2.3 Frequency-domain fatigue analysis**

For frequency-domain fatigue analysis, some researchers [2,5,6,8,16,24,25] have devised various frequency-domain models. The TB and Dirlik models are well known for their excellent performance [26-28].

**2.3.1 Dirlik model**

Dirlik [6] proposed the famous Dirlik model in 1985. The core of the model is the analytical expression between the rainflow amplitude probability density function $p_{RFC}(S)$ and four spectral moments (i.e. $\lambda_0$, $\lambda_1$, $\lambda_2$, $\lambda_4$) and is formulated by combining an exponential function and two Rayleigh functions. The expression for $p_{RFC}(S)$ is:

$$p_{RFC}(S) = \frac{1}{\sigma_X}\left[\frac{D_1}{Q}e^{-\frac{Z}{Q}} + \frac{D_2 Z}{R^2}e^{-\frac{Z^2}{2R^2}} + D_3 Z e^{-\frac{Z^2}{2}}\right] \tag{7}$$

where

$$\begin{aligned} x_m &= \frac{\lambda_1}{\lambda_0}\left(\frac{\lambda_2}{\lambda_4}\right)^{1/2}, \quad D_1 = \frac{2(x_m - \alpha_2^2)}{1+\alpha_2^2} \\ D_2 &= \frac{1-\alpha_2 - D_1 + D_1^2}{1-R}, \quad D_3 = 1 - D_1 - D_2 \\ Q &= \frac{1.25(\alpha_2 - D_3 - (D_2 R))}{D_1} \\ R &= \frac{\alpha_2 - x_m - D_1^2}{1-\alpha_2 - D_1 + D_1^2} \end{aligned} \tag{8}$$

and the normalized rainflow amplitude $Z = S/\sigma_X$. The rainflow damage per unit time $\bar{D}_{RFC}$ is expressed as [6]:

$$\bar{D}_{RFC} = \frac{v_p}{C}\sigma_X^k\left[D_1 Q^k \Gamma(1+k) + (\sqrt{2})^k \Gamma\left(1+\frac{k}{2}\right)(D_2 R^k + D_3)\right] \tag{9}$$

### 2.3.2 Tovo–Benasciutti (TB) model

Based on Rychlik's study [29], Tovo and Benasciutti [5] developed the TB model:

$$\bar{D}_{RFC} = b\bar{D}_{NB} + (1-b)\bar{D}_{RC} \tag{10}$$

where $\bar{D}_{NB}$ represents the fatigue damage per unit time calculated by the narrow-band approximation and $\bar{D}_{RC}$ represents the fatigue damage per unit time calculated by the range counting method. And the expressions for $\bar{D}_{NB}$ and $\bar{D}_{RC}$ are as follows:

$$\bar{D}_{NB} = v_0 C^{-1}\left(\sqrt{2}\sigma_X\right)^k \Gamma\left(1+\frac{k}{2}\right) \tag{11}$$

$$\bar{D}_{RC} \cong v_P C^{-1}\left(\sqrt{2}\sigma_X \alpha_2\right)^k \Gamma\left(1+\frac{k}{2}\right) = \bar{D}_{NB}\alpha_2^{k-1} \tag{12}$$

The core parameter $b$ of the TB model is expressed as [5]:

$$b = \frac{(\alpha_1 - \alpha_2)\left[1.112\left(1 + \alpha_1\alpha_2 - (\alpha_1 + \alpha_2)\right)e^{2.11\alpha_2} + (\alpha_1 - \alpha_2)\right]}{(\alpha_2 - 1)^2} \tag{13}$$

**3. Machine learning models implementation**

For the sake of convenience, this paper uses the statistics and machine learning toolbox in Matlab to develop various machine learning models for random fatigue life prediction.

**3.1 The selection of inputs and output**

In the development of a machine learning model with excellent performance, it is particularly important to select the appropriate inputs and outputs of the model. Inspired by the TB model [5], this paper chooses the core parameter $b$ of the TB model as the output for each machine learning model developed in this paper. According to several existing researches [10,12], the parameter $k$ of the *S-N* curve and the bandwidth parameters $\alpha_{0.75}$ and $\alpha_2$ are closely related to the parameter $b$ and have a deep impact on fatigue life. Therefore, $k$, $\alpha_{0.75}$ and $\alpha_2$ are selected as the inputs for each machine learning model, and their ranges are shown in Table 1.

Table 1  Ranges of the inputs and output

| Parameters | Inputs | | | Output |
|---|---|---|---|---|
| | $\alpha_{0.75}$ | $\alpha_2$ | $k$ | $b$ |
| Min | 0.42 | 0 | 3 | 0 |
| Max | 1 | 1 | 12 | 1 |

**3.2 Training dataset establishment**

Most of frequency-domain models are built by using the fatigue life calculated by time-domain fatigue analysis as standard data [2,5,6,8], and this article is no exception. Fig. 3 represents the procedure of dataset establishment for training machine learning models.

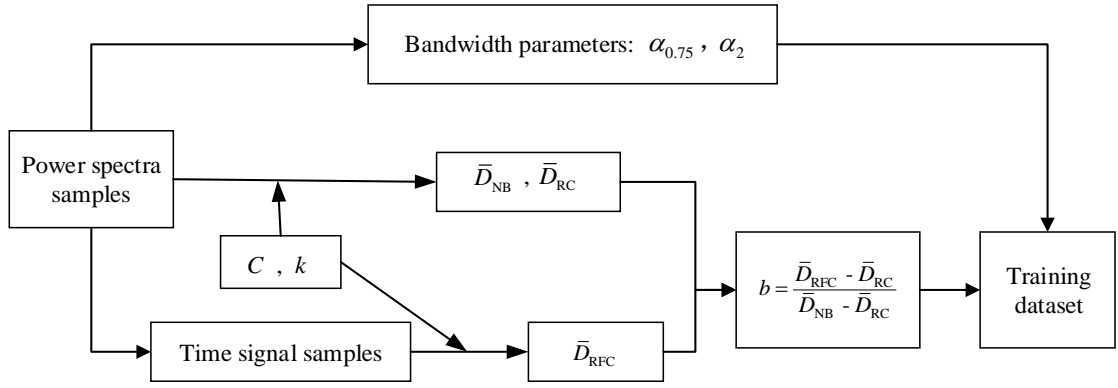

Fig. 3 The procedure of training dataset establishment

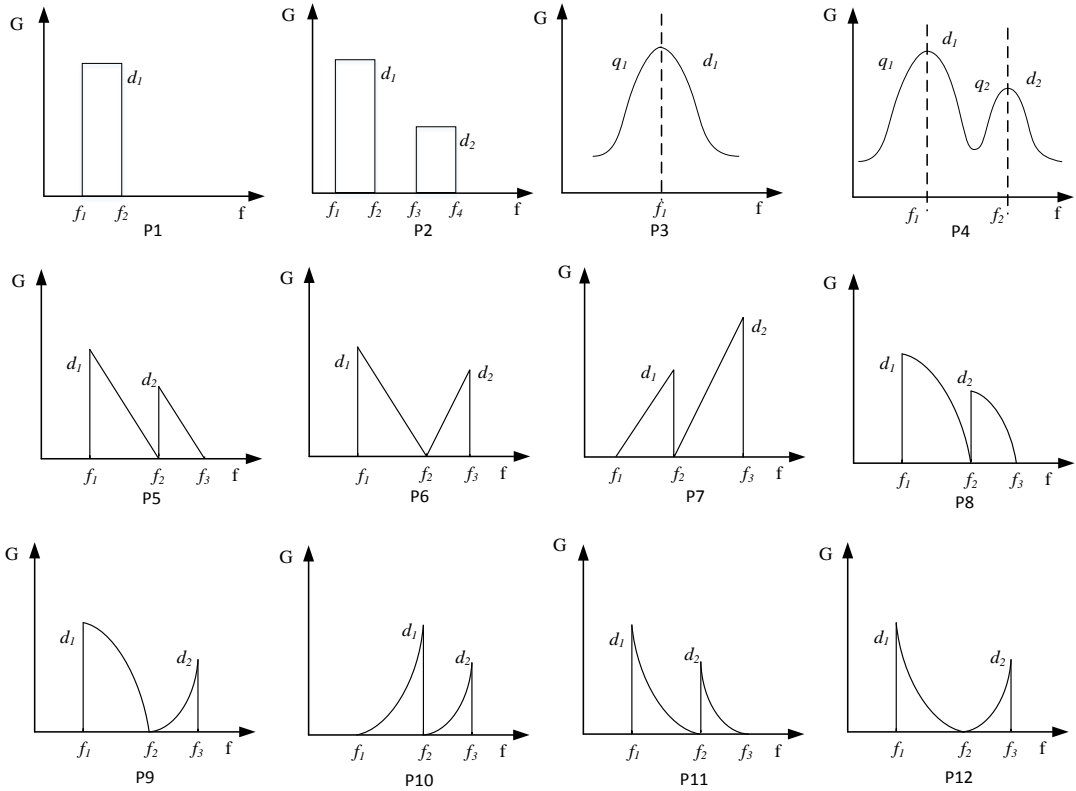

Fig. 4 Twelve different power spectra patterns used in the study

First, 12 power spectra patterns (i.e. patterns P1-P12) in Fig. 4 are used in this paper. By adjusting the spectral shape parameters (i.e. $f_1$, $f_2$, $f_3$, $d_1$, $d_2$, $q_1$, $q_2$, etc.), many power spectra samples with a wide range of bandwidth parameters $\alpha_{0.75}$ and $\alpha_2$ can be obtained. The cut-off frequency of each sample is under 300 Hz [6,15] and the range of the $0^{th}$ order spectral moment [15] is 18.6~4.13E+05 MPa$^2$. According to several existing researches [10,12], the value of $C$ can be specified as 1 in this paper, and given the parameter $k$ of the *S-N* curve, the fatigue damages $\bar{D}_{\text{NB}}$ and $\bar{D}_{\text{RC}}$ can be calculated by Eqs. (11) and (12), respectively.

Then, in order to calculate the rainflow damage $\bar{D}_{\text{RFC}}$ by time-domain fatigue analysis, the power spectra samples must be converted into time signal samples through the random amplitude and random phase method (RARP) [30]. The RARP is often expressed as follows [31,32]:

$$X(t_j) = \sum_{i=1}^{I} \left( A_i \cos \omega_i t_j + B_i \sin \omega_i t_j \right) \quad j = 0, 1 \cdots, J-1 \tag{14}$$

where

$$\begin{aligned} &E\left[A_i^2\right] = E\left[B_i^2\right] = G_X(\omega_i)\Delta\omega, \quad E\left[A_i B_i\right] = 0 \\ &\omega_i = (i-1)\Delta\omega, \quad \Delta\omega = 2\pi f_u/I, \\ &t_j = j\Delta t, \quad \Delta t = 1/f_s \end{aligned} \tag{15}$$

In Eqs. (14) and (15), $X(t_j)$ is the generated time-domain signal sample, $A_i$ and $B_i$ are two independent Gaussian random variables, $I$ is the number of discrete angular frequencies, $G_X(\omega_i)$ is the value of $G_X(\omega)$ at discrete angular frequencies $\omega_i$, $\Delta t$ is the sampling time interval, $f_u$ is the cut-off frequency of $G_X(\omega)$, $f_s$ is the sampling frequency and $J$ is the number of sampling points.

In this paper, the sampling frequency is set to 19 times of the cut-off frequency of the corresponding power spectra sample [11], the number of sampling points is ten million, and the number of discrete angular frequencies is set to 1024. As a result, the Eq. (2) holds with a maximum relative error of 5%, guaranteeing that the conversion between the time domain signals and the corresponding power spectra samples is correct.

Finally, the standard value of the parameter $b$ can be obtained by using $\bar{D}_{\text{NB}}$, $\bar{D}_{\text{RC}}$ and $\bar{D}_{\text{RFC}}$, and this paper establishes the training dataset comprising $k$, the bandwidth parameters $\alpha_{0.75}$ and $\alpha_2$ as standard input data and the parameter $b$ as standard output data.

### 3.3 Support vector machine

First proposed in 1992 [33], the support vector machine is a popular machine learning model for classification and regression. To address regression tasks, an insensitive cost function is incorporated to search for a hyperplane that minimizes the distance between individual sample and the hyperplane. Notably, the crux of SVM's efficacy resides in the utilization of the kernel function. Among the repertoire of kernel functions, prominent examples include the polynomial kernel, the Gaussian kernel, the sigmoid kernel, and the Gaussian kernel is employed in the study. Sequential Minimal Optimisation is used for the solution of SVM problems. The number of samples used to train the SVM model is 10,000, striking a balance between computational efficiency and prediction accuracy.

### 3.4 Gaussian process regression

Gaussian process regression is a nonparametric probabilistic machine learning model [34], which is on the basis of the assumption of Gaussian process prior. The Bayesian inference is usually used to solve a GPR. In the development of a GPR model, the choice of the covariance function has a significant effect on its performance. The squared exponential kernel function, one of the most commonly used covariance functions, is used to forge a GPR model in this paper. The number of samples used to train the GPR model is 1,000 after trying 500, 1,000, 2,000, 5,000 and 10,000. Besides, the hyperparameters are optimized by minimizing five-fold cross-validation loss.

### 3.5 Artificial neural network

Artificial neural network is a machine learning model that performs multiple functions such as pattern recognition, function approximation and so on. [35]. The framework of the ANN is illustrated in Fig. 5:

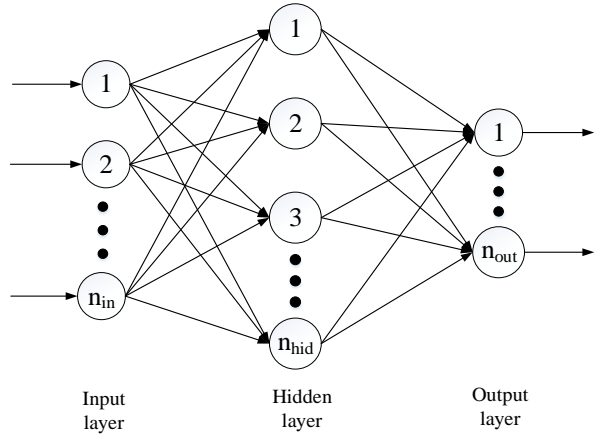

Fig. 5 The framework of an artificial neural network

Generally, a neural network whose basic element is the neuron is composed of an input layer, one or more hidden layers and an output layer. Each layer has one or more neurons that connect together to transmit information. The activation function fulfils an extremely important role in an ANN, enabling the sophisticated non-linear relationship between inputs and outputs to be modelled. And the network is trained by using the backpropagation algorithm to continually update the weight and bias of each neuron.

As stated in Section 3.1, the ANN model designed in this paper consists of three input neurons and one output neuron. To training the network, the training dataset is divided into training set, validation set, and test set with proportions of 70%, 15%, and 15%, respectively. And the other related parameter settings of the proposed ANN model are summarized in Table 2.

Table 2 Several parameter settings of the developed ANN model

| Network Parameters | Settings |
| --- | --- |
| The size of training data set | 20000 |
| The number of hidden layer | 1 |
| The number of hidden layer neurons | 7 |
| Activation function in hidden layer | Sigmoid function |
| Activation function in output layer | Linear function |
| Backpropagation algorithm | Levenberg-Marquardt algorithm |

## 4. Results and discussions

In this section, the three developed frequency-domain models (i.e. SVM, GPR and ANN models) and two classic frequency-domain models (i.e. Dirlik and TB models) are thoroughly compared in terms of prediction accuracy and computational efficiency.

**4.1 Comparison among the prediction accuracies of different models**

With the aim of evaluating the prediction accuracies of different frequency-domain models, the coefficient of determination $R^2$, the error index $EI$ and the relative error $\gamma$ are employed in the paper. The three metrics are expressed as follows, respectively:

$$R^2 = 1 - \frac{\sum(X-Y)^2}{\sum(X-\mu_X)^2} \quad (16)$$

$$EI = \sqrt{\frac{1}{n}\sum\left[\log_{10}\left(\frac{Y}{X}\right)\right]^2} \quad (17)$$

$$\gamma = \frac{Y-X}{X} \quad (18)$$

where $\mu_X$ is the mean of $X$; $X$ and $Y$ denote the fatigue damages obtained separately by the rainflow counting and by frequency-domain models (i.e. SVM, GPR, ANN, Dirlik and TB models). For the coefficient of determination $R^2$, the closer the value is to 1, the better the model; and for the error index $EI$ defined by Tovo and Benasciutti [10], the closer the value is to 0, the better the model.

**4.1.1 Performances on power spectra patterns P1 to P10**

All the three machine learning models (i.e. SVM, GPR and ANN models) developed in this paper are trained by training dataset base on power spectra patterns P1 to P10 as shown in Fig. 4. In this section, by resampling the values of $k$ and the bandwidth parameters within the ranges shown in Table 1, 10 groups [15] of test dataset base on power spectra patterns P1 to P10 are produced to compare the performances of SVM, GPR, ANN, Dirlik and TB models. Each group of test dataset is based on 1000 power spectra samples.

The $R^2$ and $EI$ for SVM, GPR, ANN, Dirlik and TB models are calculated and the results are represented in Table 3 and Table 4 respectively. In the last two rows of every table, the mean and standard deviation of the $R^2$ and $EI$ for 10 groups of test data are calculated. Fig. 6 and Fig. 7 plot the corresponding data in Table 3 and Table 4 to compare explicitly the each model's performance on every group of the test dataset.

Table 3 Comparison of the coefficients of determination $R^2$ for SVM, GPR, ANN, TB and Dirlik models

| Group | $R^2$ | | | | |
|---|---|---|---|---|---|
| | **SVM*** | **GPR*** | **ANN*** | **TB** | **Dirlik** |
| *1* | 0.9810 | 0.9797 | 0.9652 | 0.9997 | 0.9998 |
| *2* | 0.9711 | 0.9959 | 0.9737 | 0.9971 | 0.9960 |
| *3* | 0.9985 | 0.9959 | 0.9917 | 0.9912 | 0.9837 |
| *4* | 0.9979 | 0.9900 | 0.9898 | 0.9948 | 0.9882 |
| *5* | 0.9809 | 0.9735 | 0.9701 | 0.9927 | 0.9911 |
| *6* | 0.9848 | 0.9885 | 0.9743 | 0.9999 | 0.9999 |
| *7* | 0.9888 | 0.9907 | 0.9910 | 0.9850 | 0.9840 |
| *8* | 0.9879 | 0.9897 | 0.9868 | 0.9671 | 0.9692 |
| *9* | 0.9850 | 0.9922 | 0.9956 | 0.9706 | 0.9696 |
| *10* | 0.9811 | 0.9797 | 0.9652 | 0.9997 | 0.9997 |
| ***Mean*** | **0.9857** | **0.9876** | **0.9803** | **0.9898** | **0.9881** |
| ***Std*** | **0.0082** | **0.0075** | **0.0118** | **0.0120** | **0.0116** |

* the author's model

As shown in Table 3 and Fig. 6, the mean values of $R^2$ corresponding to SVM, GPR, ANN, TB and Dirlik models exhibit only minor discrepancies and this holds true for the standard deviations, indicating that the five models have almost the same performance in terms of the coefficients of determination $R^2$.

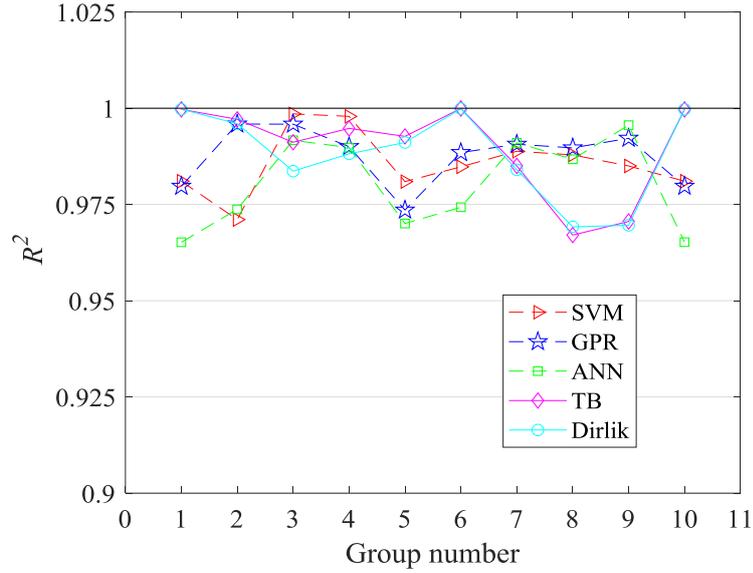

Fig. 6 Comparison of the coefficients of determination $R^2$ for SVM, GPR, ANN, TB and Dirlik models

Table 4 Comparison of the error indexes $EI$ for SVM, GPR, ANN, TB and Dirlik models

| Group | $EI$ | | | | |
|---|---|---|---|---|---|
| | SVM* | GPR* | ANN* | TB | Dirlik |
| 1 | 0.6466 | 0.8043 | 0.7391 | 2.4319 | 1.8091 |
| 2 | 0.6662 | 0.8925 | 0.7698 | 2.4452 | 1.8117 |
| 3 | 0.6448 | 0.9293 | 0.7426 | 2.4801 | 1.8516 |
| 4 | 0.6138 | 0.9174 | 0.7598 | 2.4388 | 1.7920 |
| 5 | 0.6719 | 0.8972 | 0.7900 | 2.5010 | 1.8777 |
| 6 | 0.6747 | 0.8553 | 0.7662 | 2.3711 | 1.7461 |
| 7 | 0.5957 | 0.8115 | 0.7591 | 2.4486 | 1.8651 |
| 8 | 0.5883 | 0.7821 | 0.7147 | 2.3682 | 1.7857 |
| 9 | 0.6289 | 0.8825 | 0.7585 | 2.4929 | 1.9259 |
| 10 | 0.6860 | 0.8758 | 0.7750 | 2.4392 | 1.8071 |
| **Mean** | **0.6417** | **0.8648** | **0.7575** | **2.4417** | **1.8272** |
| **Std** | **0.0342** | **0.0501** | **0.0211** | **0.0449** | **0.0525** |

* the author's model

However, according to Table 4 and Fig. 7, it is obvious that the SVM, GPR, ANN models implemented in this paper have much smaller error indexes $EI$ than the TB and Dirlik models do, which manifests that the three models implemented in this paper possess superior performances in fatigue damage predictions than the existing frequency-domain models (i.e. TB and Dirlik models) do. Furthermore, the SVM model is the best model among the SVM, GPR, ANN models because of its smallest $EI$.

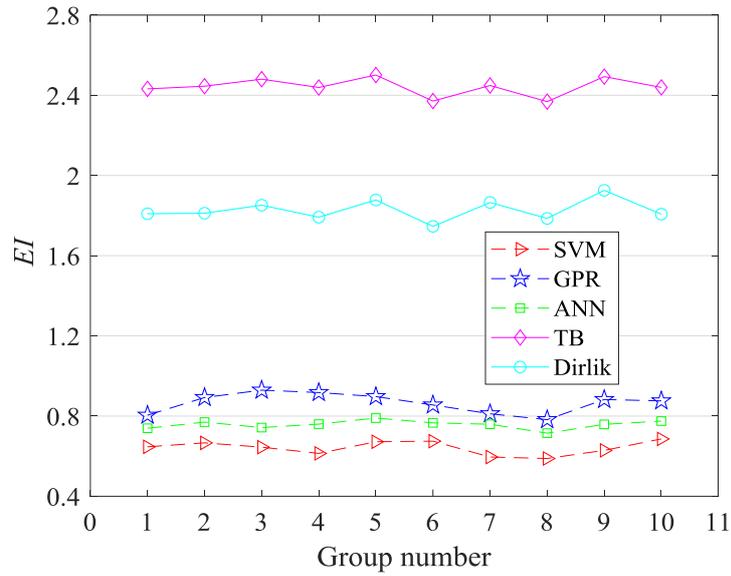

Fig. 7 Comparison of the error indexes $EI$ for SVM, GPR, ANN, TB and Dirlik models

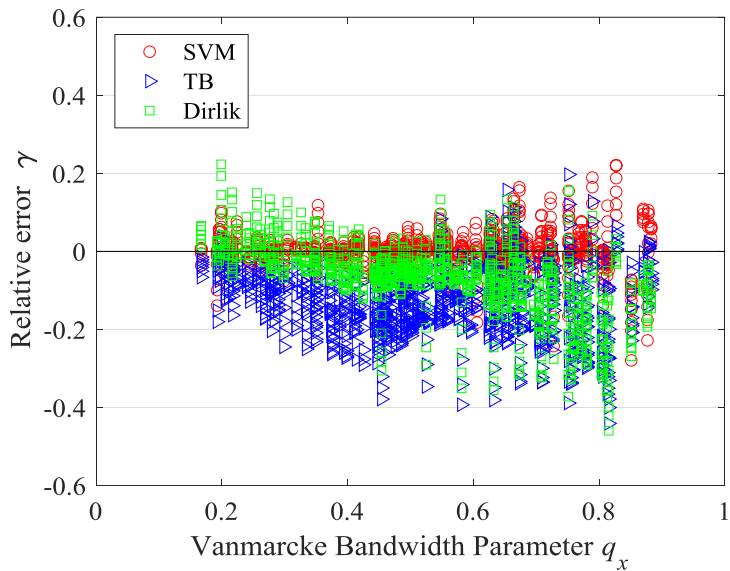

Fig. 8 Comparison of the relative errors $\gamma$ for SVM, TB and Dirlik models

Additionally, for the SVM, GPR, ANN, TB and Dirlik models, the values of the relative error $\gamma$ for Group 1 out of 10 groups of the test dataset are plotted in Fig. 8, Fig. 9 and Fig. 10. It is observed from Fig. 8 that the red points corresponding to the SVM model are more concentrated around the straight line $\gamma = 0$, suggesting that the SVM model is more accurate in predicting fatigue damages than TB and Dirlik models. Similarly, Fig. 9 and Fig. 10 show that the GPR and ANN models perform better than the TB and Dirlik models do.

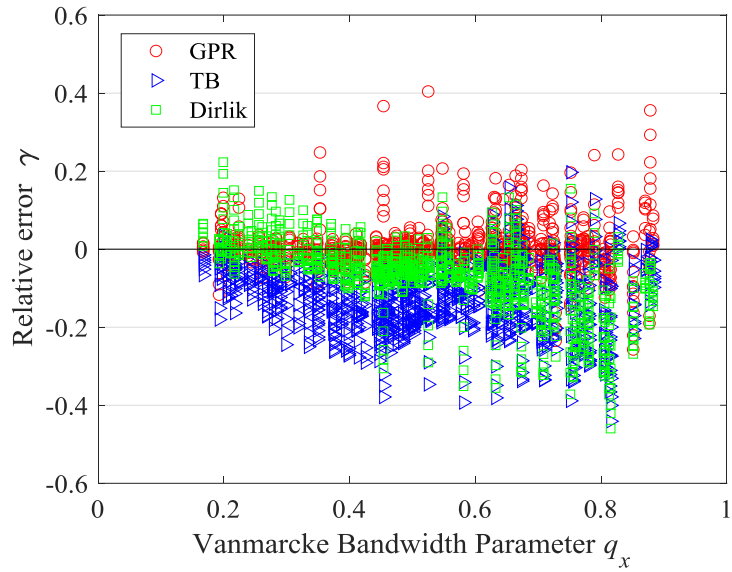

Fig. 9 Comparison of the relative errors $\gamma$ for GPR, TB and Dirlik models

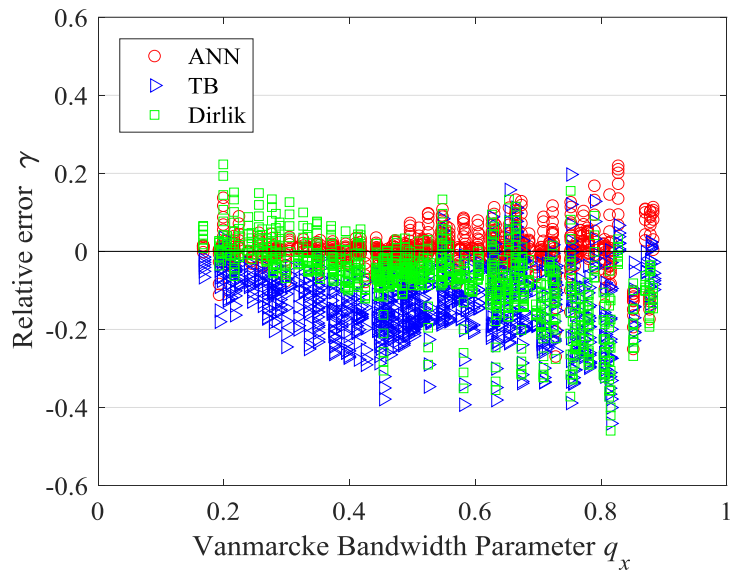

Fig. 10 Comparison of the relative errors $\gamma$ for ANN, TB and Dirlik models

### 4.1.2 Performance on unseen power spectra patterns P11 and P12

In order to verify the performance of SVM, GPR, ANN models on unseen patterns, the three models are trained on power spectra patterns P1 to P10 and are tested on unseen patterns P11 and P12 in Fig. 4. The unseen patterns P11 and P12 are used to generate a group of test dataset containing 1000 power spectra samples and the values of the relative error $\gamma$ for the test dataset are plotted in Fig. 11, Fig. 12 and Fig. 13. As can be seen from these figures, the red points denoting SVM, GPR, and ANN models cluster closely along the straight line $\gamma = 0$, demonstrating that the three models developed in this paper outperform the TB and Dirlik models in predicting fatigue damages even when tested on unseen patterns P11 and P12.

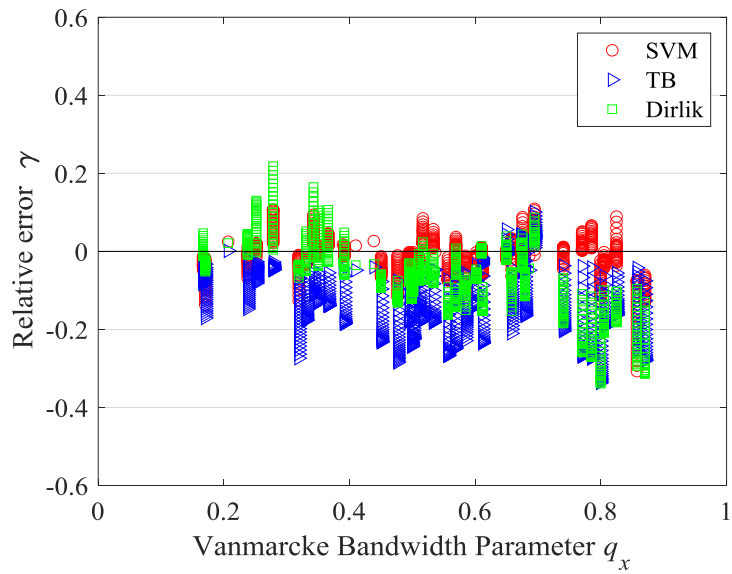

Fig. 11 Comparison of the relative errors $\gamma$ for SVM, TB and Dirlik models on unseen patterns P11 and P12

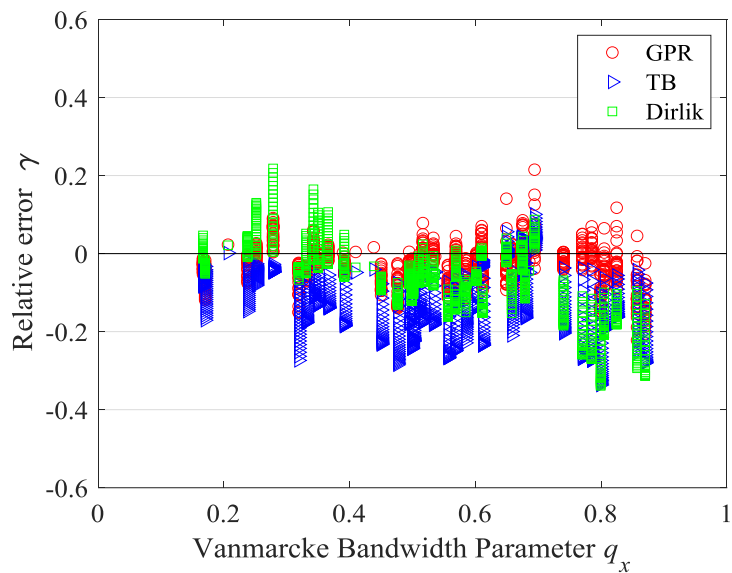

Fig. 12 Comparison of the relative errors $\gamma$ for GPR, TB and Dirlik models on unseen

patterns P11 and P12

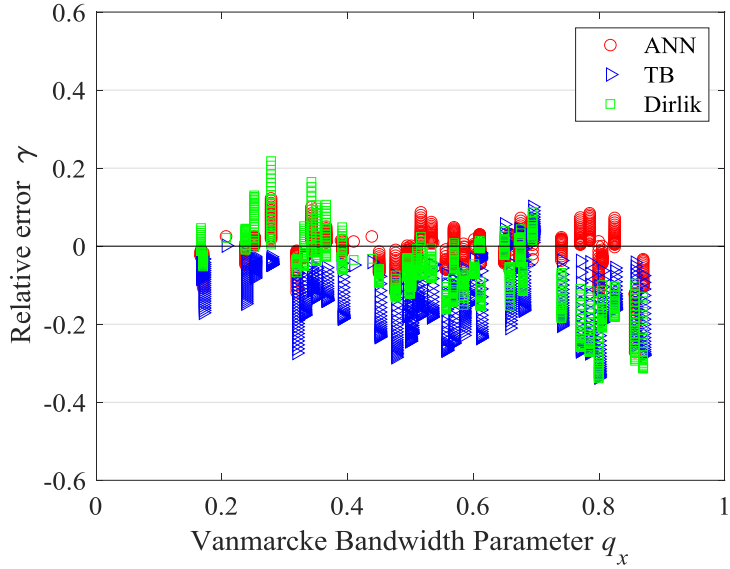

Fig. 13 Comparison of the relative errors $\gamma$ for ANN, TB and Dirlik models on unseen patterns P11 and P12

## 4.2 Comparison among different models in terms of computational efficiency

With the purpose of comparing the different models in terms of computational efficiency, the computational expenses for 1,000 power spectra samples were assessed for the frequency-domain models (i.e. SVM, GPR, ANN, TB and Dirlik models) and the time-domain model (i.e. the rainflow counting (RFC)), respectively. This assessment was carried out by means of MATLAB's tic and toc commands, and the results are tabulated in Table 5.

Table 5 Comparison of computational expenses for the different models

| Run | Run time (s) | | | | | Time |
|---|---|---|---|---|---|---|
| | Frequency | | | | | |
| | SVM* | GPR* | ANN* | TB | Dirlik | RFC |
| *1* | 21.54 | 43.36 | 4.66 | 3.49 | 3.51 | 1832.64 |
| *2* | 23.12 | 43.66 | 4.54 | 3.49 | 3.53 | 1822.63 |
| *3* | 22.93 | 44.01 | 4.50 | 3.46 | 3.43 | 1815.70 |
| *4* | 22.86 | 43.63 | 4.49 | 3.41 | 3.52 | 1823.39 |
| *5* | 23.24 | 43.59 | 4.60 | 3.50 | 3.42 | 1819.09 |
| ***Mean*** | **22.74** | **43.65** | **4.56** | **3.47** | **3.48** | **1822.69** |

*the author's model

Regarding the frequency-domain models, the run time encompasses the time of two phases: firstly, acquiring basic input parameters (i.e. bandwidth parameters) of the machine learning models through processing power spectra samples, and subsequently, employing the models to make predictions about fatigue damages. Concerning the time-domain model, the run time includes the time to perform the following several steps: converting the power spectra samples into the corresponding time domain signals through the RARP method described in Section 3.2, extracting stress ranges from these signals using the rainflow counting, and then computing fatigue damages utilizing the extracted stress ranges, material properties and Miner's rule.

According to Table 5, all the frequency-domain models have a much shorter run time than the time-domain model does, which demonstrates the advantage of the frequency-domain models in computation efficiency. Despite the fact that the SVM model is the most excellent model among the frequency-domain models in prediction accuracy ( see Section 4.1), the ANN model is much better than the SVM model and rather close to the existing TB and Dirlik models in computational efficiency. As a result, the ANN model is the best of the three developed machine learning models in terms of overall performance.

5.Conclusions

Three machine learning models, the SVM, GPR and ANN, are implemented for wideband random fatigue life predictions. The parameter $b$ of the TB model is used as the output of each model and the parameter $k$ of the *S-N* curve and the bandwidth parameters $\alpha_{0.75}$ and $\alpha_2$ of power spectra samples as the inputs. With the use of numerous power spectra samples with different bandwidth parameters and material properties related to fatigue life, the three models exhibit high generalization ability. Among all the three machine learning models, the ANN model has the best overall performance, balancing the fatigue life prediction accuracy and the computational cost. Compared with the traditional frequency-domain models: the TB and Dirlik models, the three proposed machine learning models achieve higher performance in life prediction accuracy.

**Acknowledgements**

The authors gratefully acknowledge the financial support of the Natural Science Basic Research Program of Shaanxi (Program No.2023-JC-YB-328), and the financial support of the Fundamental Research Funds for the Central Universities (Program No. ZYTS23014).